\documentclass[a4paper,12pt]{article}
\usepackage[dvips]{epsfig}

\def\at#1{}

\advance\textheight2.6cm        
\advance\topmargin-2.0cm
\advance\textwidth2.4cm
\advance\evensidemargin-1.6cm   
\advance\oddsidemargin-1.6cm

\def\phi{\varphi}
\def\eps{\varepsilon}

\begin{document}


\begin{center}

{\LARGE \bf Analytic representation of} \\~

{\LARGE \bf  critical equations of state} \\

\vspace{1cm}

\centerline{\sl {\large \bf Arnold Neumaier}}

\vspace{0.5cm}

\centerline{\sl Fakult\"at f\"ur Mathematik, Universit\"at Wien}
\centerline{\sl Oskar-Morgenstern-Platz 1, A-1090 Wien, Austria}
\centerline{\sl email: Arnold.Neumaier@univie.ac.at}
\centerline{\sl WWW: \url{http://www.mat.univie.ac.at/~neum/}}

\vspace{0.5cm}

arXiv: 1401.0291

\vspace{0.5cm}

\end{center}

\hfill \today

\vfill

\bfi{Abstract.} 
A new form is proposed for equations of state (EOS) of thermodynamic 
systems in the 3-dimensional Ising universality class.
The new EOS guarantees the correct universality and scaling behavior 
close to critical points and is formulated in terms of the scaling 
fields only -- unlike the traditional Schofield representation, which 
uses a parametric form. 

Close to a critical point, the new EOS expresses the square of the 
strong scaling field $\Sigma$ as an explicit function 
$\Sigma^2=D^{2e_{-1}}W(D^{-e_0}\Theta)$ of the thermal scaling 
field $\Theta$ and the dependent scaling field $D>0$, with a smooth, 
universal function $W$ and the universal exponents 
$e_{-1}=\delta/(\delta+1)$, $e_0=1/(2-\alpha)$.
A numerical expression for $W$ is derived, valid close to critical 
points.

As a consequence of the construction it is shown that the dependent 
scaling field can be written as an explicit function of the relevant 
scaling fields without causing strongly singular behavior of the 
thermodynamic potential in the one-phase region. 

Augmented by additional scaling correction fields, the new EOS also 
describes the state space further away from critical points. 
It is indicated how to use the new EOS to model multiphase fluid 
mixtures, in particular for vapor-liquid-liquid equilibrium (VLLE)
where the traditional revised scaling approach fails.


\newpage
\tableofcontents 

\vspace*{1cm}

\section{Introduction}\label{s.intro} 

This paper owes its existence to the desirability of a multi-phase, 
multi-component equation of state for fluids and fluid mixtures, valid 
over the whole fluid regime, with the property that, close to critical 
points, plait points, and consolute points, the correct universality 
and scaling behavior is guaranteed. 
The search for such an equation of state lead to the content of the 
present paper -- the discovery that one can write the equation of state 
of any system with a critical point in the (3-dimensional) Ising 
universality class in a natural analytic form that improves upon 
traditional formulations. 

Equations of state valid over the whole fluid range must account for 
the singular behavior of various thermodynamic observables close to
critical points (for pure fluids a liquid-vapor critical point, for 
fluid mixtures plait points at liquid-vapor equilibrium and consolute 
points at liquid-liquid equilibrium) and of 
the universal, substance independent power laws with which 
certain thermodynamic quantities scale near the critical point; 
cf. the fairly recent survey by \sca{Sengers \& Shanks} \cite{SenS}.

Because of universality, the details of the microscopic models of a 
fluid are nearly irrelevant for determining the universal features  
near the critical point; thus highly simplified models may be used.
The simplest is the lattice gas, mathematically equivalent to the 
Ising model for magnetization;\footnote{\label{f.ising}
As a result, much of the statistical mechanics literature on critical 
scaling is written in a ``magnetic'' terminology.
Most of this language is inappropriate in a fluid mixture context. 
Hence we choose here an independent notation and indicate at 
times alternative traditional notation in footnotes.
} 
see \sca{Pelissetto \& Vicari} 
\cite[(1.3)]{PelV}. One therefore says that fluids and fluid mixtures 
belong to the (3-dimensional) \bfi{Ising universality class}.\footnote{ 
For a recent verification in case of the Lennard--Jones fluid see 
\sca{Watanabe} et al. \cite{WatIH}.
} 
Initially thought to
apply to fluids composed of molecules with short range forces only,
it is now believed that ionic liquids, interacting with long range 
Coulombic forces, also belong to this universality class; see 
\sca{Gutkowski} et al. \cite{GutAS}, \sca{Schr\"oer} \cite{Schr}.
 
In the following, Section \ref{s.scaling} summarizes the background on 
critical scaling and universality in general, and special properties of 
the Ising universality class.
Section \ref{s.genPolar} introduces a generalization of the traditional
Schofield representation (\sca{Schofield} \cite{Scho}) that, close to a 
critical point, eliminates the singularities by a parameterization in 
terms of generalized polar coordinates.  
Based on this, Section \ref{s.Sigma} derives the new $\Sigma$-explicit 
EOS close to a critical point. This form of the EOS was inspired by
some explicitly solvable statistical mechanics models discussed by
\sca{Fisher \& Felderhof} \cite{FisFa,FisFb}.
Section \ref{s.Dexplicit} proves the existence of an analytic 
$D$-explicit EOS. 
Section \ref{s.explicit} extends the new $\Sigma$-explicit EOS to
states further away from the critical point, by incorporation scaling
correction fields. 
Section \ref{s.phenScal} reviews phenomenological scaling models for 
fluid mixtures, and Section \ref{s.multiphase} proposes a 
phenomenological critical scaling model for multiphase fluid mixtures.
Section \ref{s.conclusion} summarizes the main results.
The appendices provide additional material with numerical details. 

\bigskip
{\bf Acknowledgments.}
I'd like to thank Jan Stengers for pointing out an error in an 
intermediate version of this paper. I also acknowledge with pleasure 
several discussions with Ali Baharev, Waltraud Huyer, and Hermann 
Schichl on earlier versions of this manuscript, which lead to 
significant improvements.

\section{Critical scaling}\label{s.scaling} 

A \bfi{thermodynamic field} is a function of \bfi{pressure} $P$,
\bfi{absolute temperature} $T$, and the \bfi{chemical potential} $\mu_i$
of each pure component $i$ in the mixture. For a fluid with $C$ 
components, the physically realizable \bfi{thermodynamic states} form a 
$(C+1)$-dimensional manifold (with singularities) in the 
$(C+2)$-dimensional $(P,T,\mu)$-space, hence are described by an 
\bfi{equation of state} (\bfi{EOS}) relating $P$, $T$, and $\mu$.

The critical behavior of fluid (or solid) mixtures is believed to be 
characterized by the existence of two \bfi {relevant scaling 
fields},\footnote{
In traditional terminology, these are the nonlinear scaling fields.
Close to the critical point, they can be approximated by linear scaling 
fields. For the lattice gas, the linear approximations of $\Sigma$, 
$\Theta$, and the ordering field $\Omega$ introduced later
are the deviation of chemical potential, temperature, and density 
from the corresponding critical point data, while the dependent 
scaling field $D$ introduced later is approximately a linear 
combination of these and the deviation of the pressure from the 
critical point pressure.\\
More realistic fluids, and especially fluid mixtures, follow this 
pattern only roughly.
Because of the lattice gas, the locus $\Sigma=0$ is sometimes called 
the \bfi{critical isochore}, and the locus $\Theta=0$ the \bfi{critical 
isotherm}, a misleading terminology when applied to fluid mixtures.\\
Less close to the critical point, additional nonlinear correction
terms appear in all four scaling fields.
} 
the \bfi{strong scaling field} $\Sigma$ and the \bfi{thermal scaling 
field} $\Theta$. 

The strong scaling field $\Sigma$ and the thermal scaling field $\Theta$
have a clear physical meaning:
For $\Theta\le 0$, the system is in a lower density phase if $\Sigma>0$,
in a higher density phase if $\Sigma<0$, and has two coexistent lower 
and higher density phases if $\Sigma =0$.  At $\Theta<0=\Sigma$, we 
have a \bfi{first-order phase transition}, i.e., some thermodynamic 
response functions possess a jump discontinuity.
The inequality $\Theta>0$ defines a low density part of the phase space,
connected smoothly to both phases at $\Theta<0$.
Near the vapor-liquid critical point, the condition $\Theta>0>\Sigma$ 
approximately corresponds to the conventional definition of 
supercritical, which in an engineering context means that both pressure 
and temperature are above the pressure and temperature of the critical 
point. Critical points are characterized by $\Sigma =\Theta=0$. 
Generically, for fluid mixtures with $C$ components, they form a 
\bfi{critical manifold} of dimension $C-1$ (\sca{Griffiths} \cite{Gri}) 
in the $(C+1)$-dimensional thermodynamic state space of the fluid. 

The form of any EOS valid close to a critical point is strongly 
restricted by renormalization group arguments from statistical 
mechanics; see, e.g., \sca{Fisher} \cite{Fis}, \sca{Zinn-Justin} 
\cite{Zin}. These give rise to various scaling laws
for particular thermodynamic variables, as pressure, temperature, and 
chemical potential approach a critical point along specific 
trajectories, which must be reproduced by any accurate EOS.

The origin of the renormalization group and hence of the scaling laws 
is the fact that in a microscopic description of a macroscopic system, 
the mesoscopic length scale on which the thermal averaging is done can 
be changed without affecting the thermodynamic limit. 
In general, the renormalization group applies to an infinite number of 
scaling fields, of which only very few are relevant close to
a critical point; their number equals the number of degrees of freedom
near an isolated critical point. Since we can find isolated critical 
points in many mixtures at fixed composition, where thermodynamic 
states have only two degrees of freedom, there are only 
two\footnote{\label{f.tricrit}
If three or more near-critical phases coexist, there may be a nearby 
tricritical or multicritical point (\sca{Griffiths \& Widom} 
\cite{GriW}, \sca{Hankey} et al. \cite{HanCS}, 
\sca{Griffiths} \cite{Gri}, \sca{Mistura} \cite{Mis}), a situation not 
covered by the Ising universality class. 
An analysis of this situation requires further research; cf. footnote
${}^{\ref{f.multicrit}}$ below.
} 
relevant scaling fields $\Sigma$ and $\Theta$ in mixtures. 
The infinitely many 
remaining \bfi{scaling correction fields} $I_1,I_2,\dots$ determine 
deviations from the power laws.\footnote{
Traditionally, the scaling correction fields were called 
\bfi{irrelevant scaling fields}, but at the presently available 
accuracies they are far from irrelevant numerically.
The notation used in the literature for the scaling fields is not 
uniform. The notation used in \sca{Fisher} \cite{Fis} is related to the 
present one by $f=D$, $h_1=\Theta$, $h_2=\Sigma$, $h_{k+2}=I_k$. 
The notation used in \sca{Sengers \& Shanks} \cite{SenS} corresponds to 
$h_1=\Sigma$, $h_2=\Theta$, $h_3=D$.
} 

The existence of the renormalization group implies\footnote{
The traditional proofs guarantee the scaling law in the form of an 
asymptotic series only; so one expects the scaling law to hold at least
close to the critical point. However, in the exactly solvable models
of \sca{Fisher \& Felderhof} \cite{FisFa,FisFb} and 
\sca{Reuter \& Bugaev} \cite{ReuB}, the scaling law can be seen to be 
valid globally.
} 
that there is a scaling equation of state
\lbeq{e.D}
D(P,T,\mu) = S(\Sigma,\Theta,I_1,I_2,\dots)
\eeq
with a three times continuously differentiable \bfi{dependent scaling 
field} $D=D(P,T,\mu)$ and an (except for $\Sigma=0>\Theta$) three times 
continuously differentiable \bfi{scaling function} $S$ satisfying the 
exact \bfi{scaling relation}
\lbeq{e.scaling}
\lambda S(\Sigma,\Theta,I_1,I_2,\dots)
 = S(\lambda^{e_{-1}}\Sigma,\lambda^{e_0}\Theta,
      \lambda^{e_1}I_1,\lambda^{e_2}I_2,\dots)
    \Forall \lambda>0,
\eeq
where 
\lbeq{e.critEx}
1>e_{-1}>e_0>0>e_1 \ge e_2 \ge \ldots
\eeq
are \bfi{universal} (i.e., substance-independent) 
\bfi{critical exponents}.
In a \bfi{classical EOS}, $D$, $\Theta$, and $\Sigma$ (or
thermodynamic quantities derived from this, such as an order parameter) 
are related by an analytic nonlinear equation without any singularities.
It is well-known that this leads to critical points with the same 
exponents as for the van der Waals EOS (in the present notation 
$e_0=\half$ and $e_{-1}=\frac{3}{4}$),
corresponding to the mean field approximation in statistical mechanics.
These exponents match neither experimental data close to the critical 
point nor theoretical predictions from Ising-like models, whose 
universal features are valid for the whole Ising universality class. 
Accurate numerical values for the most important critical 
exponents\footnote{
There is a multitude of related critical exponents for various 
critical scaling relations between particular thermodynamic observables;
see, e.g., \sca{Pelissetto \& Vicari} \cite{PelV}.
In dimension $d=3$, these are given with their traditional label by 
\vspace{-0.3cm}
\[
\bary{c}
\alpha:=2-1/e_0,~~~\beta:=(1-e_{-1})/e_0,~~~
\gamma:=(2e_{-1}-1)/e_0,~~~\delta:=e_{-1}/(1-e_{-1}),\\
\nu:=1/3e_0,~~~\eta:=5-6e_{-1},~~~\omega:=-3e_1,~~~
y_m:=3e_m,~~~\Delta_m:=-e_m/e_0.
\eary
\]
\vspace{-0.4cm}

\noindent
The currently best values, taken from \sca{Hasenbusch} \cite{Has.Fin}
and \sca{Newman \& Riedel} \cite[Table V]{NewR} are
\vspace{-0.2cm}
\[
\nu=0.63002(10),~~~ \eta=0.03627(10),~~~ \omega=0.832(6),~~~
\Delta_2=0.98(6),~~~\Delta_3=1.07(11).
\]
\vspace{-0.6cm}

\noindent
Using the propagation formulas 
$\sigma_{f(x)}\approx |f'(\mu_x)|\sigma_x$ 
for the standard deviation $\sigma_x$ of a random variable $x$ with 
mean $\mu_x$, this yields the above values for $e_{-1}=(5-\eta)/6$, 
$e_0=1/3\nu$, $e_1=-\omega/3$, and $e_m=y_m/3$ ($m=2,3$).
} 
are 
{\small
\lbeq{e.ae}
e_{-1} = 0.82729(2),~
e_0 = 0.52908(9),~
e_1 = -0.277(2),~
e_2= -0.56(3),~
e_3= -0.61(6)
\eeq
} 
where the numbers in parentheses denote one standard deviation 
uncertainty per unit the last place. 
In particular, the EOS for every real fluid or fluid mixture is 
nonclassical and intrinsically nonanalytic near critical points. 

All general universality and scaling properties are consequences of 
the scaling relation \gzit{e.scaling} for \gzit{e.D} and the known 
smoothness properties. In particular, both renormalization group theory 
and experimental evidence show that once the scaling fields are 
normalized to get rid of multipliers in their definition that do not 
affect the validity of the scaling relation, $S$ is a universal 
function of its arguments. 

We now exploit special properties of the 3-dimensional Ising model. 
The first, well-known property is the fact that the Ising model has a 
reflection symmetry, which implies that the scaling function $S$ in 
\gzit{e.D} is an even function of $\Sigma$, hence depends on $\Sigma$ 
only through $\Sigma^2$,
\lbeq{e.sym}
S(\Sigma,\Theta,I_1,I_2,\dots) = s(\Sigma^2,\Theta,I_1,I_2,\dots).
\eeq
Since $S$ is a universal function, this holds generally in the Ising 
universality class. Thus, as far as fluids and fluid mixtures are 
described by the Ising universality class, they inherit this symmetry 
when the scaling fields are properly chosen. 

The second property is a hitherto apparently unnoticed fact: 
For the 3-dimensional Ising model, and hence in all models from the 
3-dimensional Ising universality class, the dependent scaling field $D$,
which must vanish at the critical point, is positive in a punctured 
neighborhood of the critical point. 
This observation is crucial for the derivation of the new analytic 
representation of the critical EOS, as it allows us to make in 
\gzit{e.scaling} the special choice $\lambda:=D^{-1}$. We obtain
\[
S(D^{-e_{-1}}\Sigma,D^{-e_0}\Theta,
                              D^{-e_1}I_1,D^{-e_2}I_2,\dots)
  =D^{-1} S(\Sigma,\Theta,I_1,I_2,\dots)  \Forall \lambda>0.
\]
In view of \gzit{e.sym}, the EOS \gzit{e.D} takes the implicit form
\lbeq{e.iEOS0}
s(D^{-2e_{-1}}\Sigma^2,D^{-e_0}\Theta,
                  D^{-e_1}I_1,D^{-e_2}I_2,\dots)=1.
\eeq
Remarkably, the thermodynamic state space defined by an implicit 
equation of this form is, independent of the particular choice of the 
function $s$, automatically invariant under the scaling transformation
\lbeq{e.scTrans}
\Sigma\to \lambda^{e_{-1}}\Sigma,~~~
\Theta\to \lambda^{e_0} \Theta,~~~
D \to \lambda D,~~~
I_k \to \lambda^{e_k}I_k
\eeq
with arbitrary $\lambda>0$. Hence \gzit{e.iEOS0} is a full embodiment 
of all universality and critical scaling properties,  even when $s$ 
itself satisfies no scaling relation. 
Of course, the function $s$ must be such that, at least near a critical 
point, ome may solve \gzit{e.iEOS0} for $D$ when 
$\Sigma,\Theta,I_1,I_2,\dots$ are given.
Should there be more than one solution, thermodynamic stability
considerations\footnote{
For two phases of a pure substance, $D$, $\Theta$, and $\Sigma$ are the 
renormalized analogue of pressure, temperature, and molar Gibbs free 
energy (which for pure substances equals the chemical potential). 
Therefore the stable phase at fixed $\Theta$ consists of the branches 
with the smallest value of $\Sigma$ at fixed $D$. Drawing a $(D,\Sigma)$
diagram then shows that $D$ has the largest value at fixed $\Sigma$.
} 
imply that the largest solution describes the stable phase.
In case of ties, several phases coexist; cf. \sca{Neumaier} 
\cite{Neu.critEOS}.

Very close to the critical point, the scaling correction 
fields can be neglected, and \gzit{e.D} simplifies to 
\lbeq{e.critSimple}
D = S^\crit(\Sigma,\Theta),
\eeq
where 
\lbeq{e.deltacr}
S^\crit(\Sigma,\Theta):=S(\Sigma,\Theta,0,0,\ldots).
\eeq
Similarly, \gzit{e.iEOS0} simplifies to
\lbeq{e.iEOScrit}
s^\crit(D^{-2e_{-1}}\Sigma^2,D^{-e_0}\Theta)=1,
\eeq
where $s^\crit$ is an analytic function but need not satisfy a scaling 
relation.

However, unless both $\Sigma$ and $\Theta$ are tiny, the influence of 
(at least) the first scaling correction field is significant and leads 
to so-called \bfi{corrections to scaling} (\sca{Wegner} \cite{Weg}).

\section{Order parameter and generalized polar coordinates}
\label{s.genPolar}

\def\xmax{\xi_{\max}}
\def\tmax{\theta_{\max}}
\def\coex{{\fns{coex}}}

Traditionally, the properties close to the critical point (where all 
scaling correction fields $I_k$ may be neglected) are discussed 
in terms of a thermodynamic relations between $|\Sigma|$, $\Theta$, 
and the \bfi{asymptotic order parameter}\footnote{
For the statistical mechanics treatment of the Ising model, the 
appropriate language uses a ``magnetic'' terminology. There, 
in the immediate neighborhood of the critical point, the thermal 
scaling field $\Theta$ is proportional to $T-T^\crit$, where $T^\crit$ 
is the temperature at the critical point, the strong scaling field 
$\Sigma$ is proportional to the magnetic field $H$, the order parameter 
$\Omega$ is proportional to the magnetization $M$, and the dependent
scaling field $D$ is proportional to the Gibbs free energy 
$V^{-1}\log Z$ minus an analytic function of $T$.
(Much of the discussion in the literature simply equates 
$\Theta=T-T^\crit$, $\Sigma=H$, and $\Omega=M$.)
\\
The proportionality factors are the only nonuniversal 
features in the relations between $\Sigma$, $\Theta$, $\Omega$, and $D$.
Further away from the critical point, additional nonlinear terms are 
needed: Thus 
$D$ is $V^{-1}\log Z$ minus a function of $T-T_c$ and $H$, even in $H$,
$\Theta$ is a function of $T-T_c$ and $H$, even in $H$, and 
$\Sigma$ is a function of $T-T_c$ and $H$, odd in $H$. 
In this case, one must also account for the scaling correction fields.
} 
\lbeq{e.Omega}
\Omega:=\Big(\frac{dD}{d\Sigma}\Big)_\Theta 
= \frac{d}{d\Sigma} S^\crit(\Sigma,\Theta).
\eeq
Following \sca{Schofield} \cite{Scho}, this is done by a 
parameterization in terms of variables adapted to the geometry near a 
critical point, chosen to remove the singularity at the critical point. 
In a parametric representation, the relevant fields are expressed in 
terms of a radial parameter $r\ge 0$, a measure of distance from the 
critical point, and an angular parameter 
$\phi \in [-\phi_{\max{}},\phi_{\max{}}]$, taking the values 
$\phi=\pm \phi_{\max{}}$ at the two sides of the coexistence curve.
Close to the critical point, asymptotic scaling properties hold for 
any parameterization in terms of \bfi{generalized polar coordinates} 
of the form 
\lbeq{e.genPolar}
D=r^{e_D} c_D(\phi^2),~~~
\Sigma=r^{e_\Sigma}\phi\, c_\Sigma(\phi^2),~~~
\Theta=r^{e_\Theta} c_\Theta(\phi^2),~~~
\Omega=r^{e_\Omega} \phi\, c_\Omega(\phi^2)
\eeq
for $r\ge 0$ and $|\phi|\le \phi_{\max{}}$,
with appropriate critical exponents $e_D$, $e_\Theta$, $e_\Sigma$, 
$e_\Omega$ and analytic functions $c_D\ge 0$, $c_\Theta$, 
$c_\Sigma\ge 0$, and $c_\Omega\ge 0$ of the single variable 
\lbeq{e.xi}
\xi:=\phi^2 \in [0,\xi_{\max{}}],~~~\xi_{\max{}}:=\phi_{\max{}}^2.
\eeq
The particular form \gzit{e.genPolar} of the parameterization is chosen 
to reflect the reflection symmetry of the Ising model and the fact that 
for $\Sigma\to\pm\,0$, the order parameter $\Omega$ vanishes if 
$\Theta\ge 0$, whereas for $\Theta<0$, it has the sign of $\Sigma$ and 
grows with the distance from the critical point.

In order to be equivalent to \gzit{e.critSimple}, the parameterizing 
functions $c_D$, $c_\Theta$, $c_\Sigma$, and $c_\Omega$ must, in 
addition to the above sign constraints, satisfy some simple conditions 
that guarantee unique coordinates for the $(\Theta,\Sigma)$-plane:
For $\xi\in[0,\xi_{\max{}}]$, $c_\Sigma$ must vanish at 
$\xi=\xi_{\max{}}$, and $c_\Theta$ must vanish at $\xi=\xi_0$ for some
$\xi_0$ with $0<\xi_0<\xi_{\max{}}$. Moreover, 
$|\Sigma|^{-e_\Theta/e_\Sigma}\Theta$ must take every real value just 
once. This requires that 
\[
Y(\xi)
:=\frac{c_\Theta(\xi)}{(\sqrt{\xi}c_\Sigma(\xi))^{e_\Theta/e_\Sigma}},
\]
the analogue of the cotangent function for ordinary polar coordinates, 
must be strictly monotone decreasing.

The invariance under the scaling relations \gzit{e.scTrans} is ensured 
for the critical exponents 
\lbeq{e.eNew}
e_D=1,~~~e_\Sigma=e_{-1},~~~e_\Theta=e_0,~~~e_\Omega=1-e_{-1}.
\eeq
But the critical exponents are determined only up to a common factor, 
as the reparameterization $r'=r^k$ shows. Their ratios are universal 
and determine the traditional critical exponents
\lbeq{e.betaDelta}
\beta:=\frac{e_\Omega}{e_\Theta}=\frac{1-e_{-1}}{e_0},~~~
\delta:=\frac{e_\Sigma}{e_\Omega}=\frac{e_{-1}}{1-e_{-1}},
\eeq
\[
2-\alpha:=\frac{e_D}{e_\Theta}=\frac{1}{e_0}=\beta(\delta+1),
\]
in terms of which
\lbeq{e.em10}
e_{-1}=\frac{\delta}{\delta+1},~~~
e_0=\frac{1}{2-\alpha}=\frac{1}{\beta(\delta+1)}.
\eeq
Without loss of generality, the free factor may be chosen to make 
$e_\Theta=1$. If this choice is made, the most general setting given 
above reduces to the situation 
\lbeq{e.eTrad}
e_D=2-\alpha=\beta(\delta+1),~~~e_\Sigma=\beta\delta,~~~
e_\Theta=1,~~~e_\Omega=\beta,
\eeq
extensively discussed in \sca{Pelissetto \& Vicari} \cite{PelV}.
The relations are spelled out in more details in Appendix 
\ref{s.uniGen}.

At fixed $\Theta\ne 0$, we have for all $\phi>0$
\[
r=\Big(\frac{\Theta}{c_\Theta(\phi^2)}\Big)^{1/e_\Theta}, 
\]
\[
D=|\Theta|^{\beta(\delta+1)}\ 
          \frac{ c_D(\phi^2)}{ |c_\Theta(\phi^2)|^{\beta(\delta+1)}},~~~
\Sigma=|\Theta|^{\beta\delta}\ 
       \frac{\phi c_\Sigma(\phi^2)}{|c_\Theta(\phi^2)|^{\beta\delta}},
\]
hence (writing $c'(\xi):=dc(\xi)/d\xi$)
\[
\bary{lll}
\D\Big(\frac{dD}{d\phi}\Big)_\Theta
&=&\D|\Theta|^{\beta(\delta+1)}\ 
   \frac{c_D'2\phi |c_\Theta|^{\beta(\delta+1)}
   -c_D\beta(\delta+1)|c_\Theta|^{\beta(\delta+1)-1}
       |c_\Theta|'2\phi}{|c_\Theta|^{2\beta(\delta+1)}}\\[4mm]
&=&\D|\Theta|^{\beta(\delta+1)}\ 
   \frac{2\phi( c_D' |c_\Theta|-\beta(\delta+1) c_D |c_\Theta|')}
        {|c_\Theta|^{\beta(\delta+1)+1}},
\eary
\]
\[
\bary{lll}
\D\Big(\frac{d\Sigma}{d\phi}\Big)_\Theta
&=&\D|\Theta|^{\beta\delta}\ 
   \frac{(c_\Sigma+\phi c_\Sigma'2\phi) |c_\Theta|^{\beta\delta}
    -\phi c_\Sigma\beta\delta|c_\Theta|^{\beta\delta-1}|c_\Theta|'2\phi}
        {|c_\Theta|^{2\beta\delta}}\\[4mm]
&=&\D|\Theta|^{\beta\delta}\ 
   \frac{c_\Sigma |c_\Theta|
       +2\phi^2(c_\Sigma' |c_\Theta|- \beta\delta c_\Sigma |c_\Theta|')}
        {|c_\Theta|^{\beta\delta+1}}.
\eary
\]
Therefore
\[
\Big(\frac{dD}{d\Sigma}\Big)_\Theta =
|\Theta|^\beta \frac{\phi c_\Omega(\phi^2)}{ |c_\Theta|(\phi^2)^\beta}
=\Omega,
\]
where
\lbeq{e.efOmega}
c_\Omega=\frac{c_D' |c_\Theta|-\beta(\delta+1) c_D |c_\Theta|'}
      {\half c_\Sigma |c_\Theta|
       +\xi(c_\Sigma'|c_\Theta| -\beta\delta c_\Sigma |c_\Theta|')}
=\frac{c_D' c_\Theta-\beta(\delta+1) c_D c_\Theta'}
      {\half c_\Sigma c_\Theta
       +\xi(c_\Sigma'c_\Theta -\beta\delta c_\Sigma c_\Theta')}.
\eeq
By reflection symmetry, the same result also holds for $\phi<0$, and
hence generally.

The parameterizing functions $c_D$, $c_\Theta$, $c_\Sigma$, and 
$c_\Omega$ determine the precise form of the implicit dependence of $D$,
$\Theta$, $\Sigma$, and $\Omega$ defined by \gzit{e.genPolar}.
Two of the parameterizing functions may be chosen fairly freely and 
determine the coordinate transformation;
the remaining ones are then fixed by the relation \gzit{e.efOmega} and 
the universal EOS. The initially unknown universal part is 
determined numerically from fits to data from experiment or from 
perturbative microscopic calculations or statistical simulation of a
statistical mechanics model from the Ising universality class.

\section{A $\Sigma$-explicit critical EOS}\label{s.Sigma}

Continuing the discussion of Section \ref{s.genPolar}, we now discuss 
in more details numerical results for a special choice of generalized 
polar coordinates. We use the critical exponents \gzit{e.eNew} and 
define coordinates through 
\lbeq{e.fNew}
c_D(\xi)=1,~~~c_\Sigma(\xi)=C(\tmax-\xi),~~~
c_\Theta(\xi)=\tmax-\xi,~~~c_\Omega(\xi)=B(\tmax-\xi)^{-1}
\eeq
with analytic functions $C(\theta)=C^\crit(\theta)$ and 
$B(\theta)=B^\crit(\theta)$ of 
\[
\theta:=\tmax-\xi.
\]
From \gzit{e.genPolar}, \gzit{e.eNew}, and \gzit{e.fNew}, we get the
explicit coordinate definitions
\lbeq{e.cooNew}
r:=D,~~~\theta:=D^{-e_0}\Theta,~~~\phi:=\sign \Sigma \sqrt{\tmax-\theta}
\eeq
and the \bfi{asymptotic scaling EOS}
\lbeq{e.asEOS}
\Sigma^2=D^{2e_{-1}}W^\crit(D^{-e_0}\Theta),~~~
\Omega^2=D^{2(1-e_{-1})}\Delta^\crit(D^{-e_0}\Theta),~~~
\sign \Omega=\sign\Sigma,
\eeq
with analytic
\lbeq{e.Ccrit}
W^\crit(\theta)=(\tmax-\theta)C^\crit(\theta)^2,~~~
\Delta^\crit(\theta)=(\tmax-\theta)B^\crit(\theta)^{-2}.
\eeq
\gzit{e.asEOS} captures the asymptotic critical behavior in a simple 
and comprehensive way.
\gzit{e.asEOS} is explicit in $\Sigma^2$, and has the form 
\gzit{e.iEOScrit} with
\[
s^\crit(\sigma,\theta)=1+\sigma^2-W^\crit(\theta),~~~
\sigma:=D^{-e_{-1}}\Sigma.
\]
In particular, $W^\crit$ can be obtained from the scaling function 
$S$ in \gzit{e.sym} by solving \gzit{e.D} for $\Sigma^2$. 
This proves that $W^\crit$ has the same differentiability 
properties as the scaling function $S$. 
In the literature, this scaling function is generally assumed to be 
analytic except for a possible singularity at $\Theta=0$.
This is based on arguments of \sca{Griffiths} \cite{Gri.an}
who suggested what is now called \bfi{Griffiths analyticity}, that 
thermodynamic functions should be analytic except at the critical point.
We show in Appendix \ref{s.uniGen} that assuming $W^\crit$ to be 
analytic has the same consequences as Griffiths analyticity for the 
high-temperature and low temperature expansions that fully capture the
critical behavior.

\begin{figure}[t!]
\caption{The asymptotic universal functions $C(\theta)$, $B(\theta)$, 
and $W(\theta)$, and the ``linear'' approximation (dashed) from 
\gzit{e.GammaLin}.}
\label{f.criticalG}
\begin{center}
    \psfig{figure=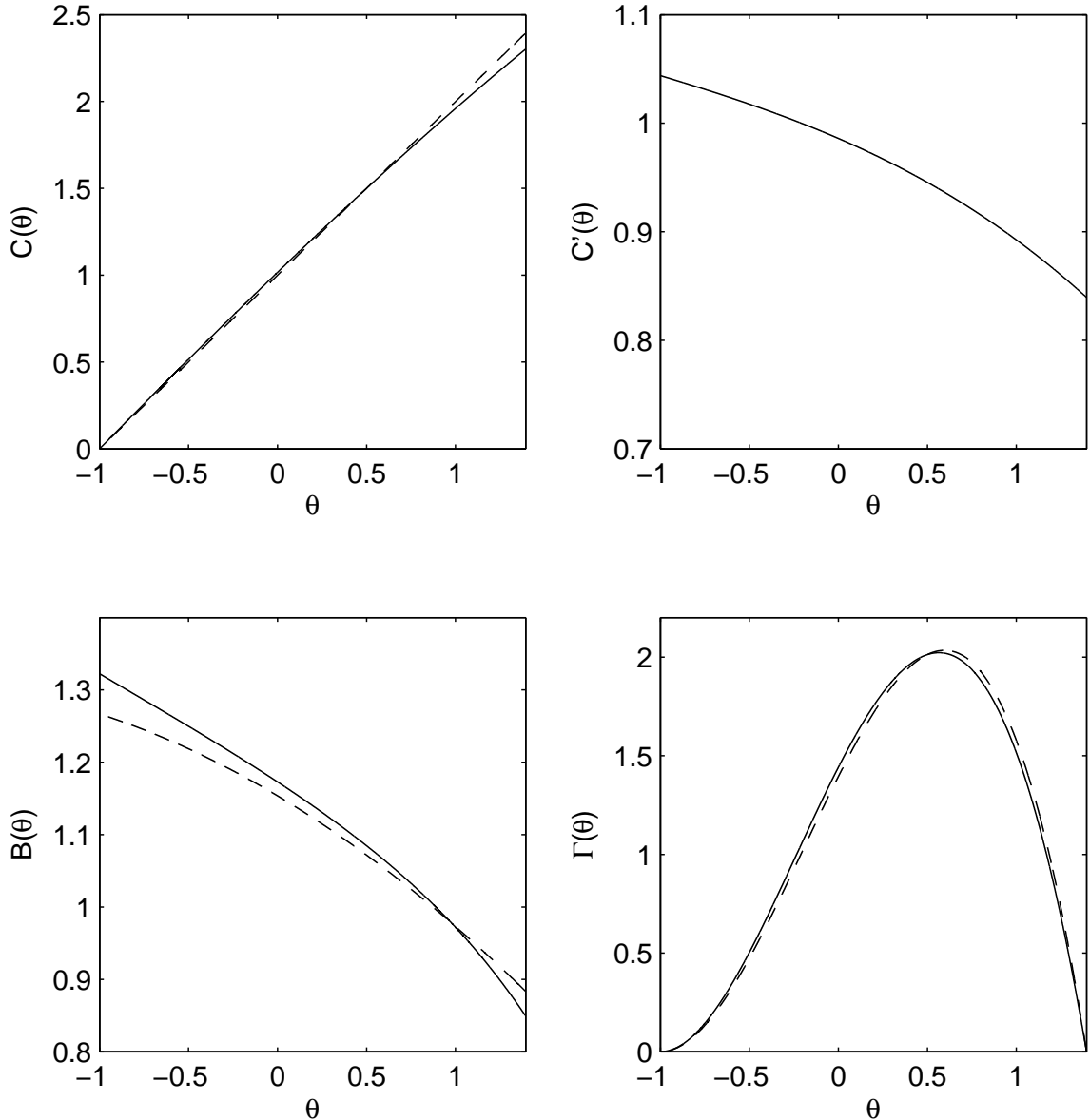,height=16cm} 
\end{center}
\end{figure}

The parameter $\theta$ takes values only in a finite interval 
$[\theta_\coex,\tmax]$, where
\[
\theta_\coex:=\tmax-\xmax<0,
\]
and we have $\theta=\theta_\coex$ along the coexistence manifold. 
The asymptotic universal function $W^\crit(\theta)$
is positive for $\theta_\coex<\theta<\tmax$ and vanishes at 
both end points, with a double zero at $\theta=\theta_\coex$.
Thus we have \gzit{e.Ccrit} with a real analytic, nonnegative function 
$C^\crit$ of $\theta\in[\theta_\coex,\theta_{\max}]$.
We may normalize the thermal scaling field such that 
$\theta_\coex=-1$; then $\tmax$ is a universal constant, and we may 
normalize the strong scaling field such that 
\lbeq{e.gcrit}
C^\crit(-1)=0,~~~C^\crit(0.5)=1.5.
\eeq
The normalization by $C^\crit(0.5)=1.5$ is a convenient choice 
satisfied for
\lbeq{e.GammaLin}
C^\crit_\fns{lin}(\theta)=1+\theta,
\eeq 
which (cf. Figure \ref{f.criticalG}) already gives a reasonable first 
approximation $W^\crit_\fns{lin}(\theta)$ to 
$W^\crit(\theta)$.

Since $\Sigma=0$ is equivalent to $W^\crit(\theta)=0$ and 
$\Theta<0$ on the coexistence manifold, we must have there 
$C^\crit(\theta)=0$, hence $\theta=-1$. Therefore the equation for the 
coexistence manifold simply becomes
\lbeq{e.asCoex}
\Theta(P,T,\mu)=-D(P,T,\mu)^{e_0}.
\eeq
\at{Reconcile the two phase description with Lee--Yang phase transition 
theory. There the two analytic functions on both sides are quite 
different, not (as in van der Waals) related by an algebraic equation.
Here would belong the metastable discussion.}

From \gzit{e.efOmega}, \gzit{e.fNew}, and \gzit{e.em10} we deduce the 
formula
\[
B(\theta)=\frac{e_0\theta}{2}C(\theta)
       +(\tmax-\theta)\Big(e_{-1}C(\theta)-e_0\theta C'(\theta)\Big).
\]
With the normalization \gzit{e.gcrit}, $C(\theta)$ and hence 
$W^\crit(\theta)$ and $B(\theta)$ are universal analytic functions 
of $\theta\in[-1,\tmax]$, and we have $C(\theta)>0$ for 
$-1<\theta<\tmax$.

The detailed information about the form of the universal function 
$W^\crit$ just spelled out may be derived from computations based 
on microscopic models of a fluid. Using information from
\sca{Butera \& Pernici} \cite{ButP} (see Appendix \ref{s.num} for more 
details) we find the approximation
\lbeq{e.Gfit}
C^\crit(\theta)=(\theta+1)\Big(1-(\theta-0.5)c(\theta)\Big),
\eeq
\lbeq{e.Gfit1}
c(\theta)=0.03339+0.005356\theta+0.001185\theta^2,~~~\tmax=1.39444.
\eeq
Figure \ref{f.criticalG} displays $W^\crit$, $C^\crit(\theta)$, 
$B^\crit(\theta)$ for this choice and for the simpler approximation 
\gzit{e.GammaLin}.

\section{A $D$-explicit formulation}\label{s.Dexplicit}

\begin{figure}[t!]
\caption{The asymptotic universal function $r(Q)$ for $a=1$.}
\label{f.criticalQ}
\begin{center}
    \psfig{figure=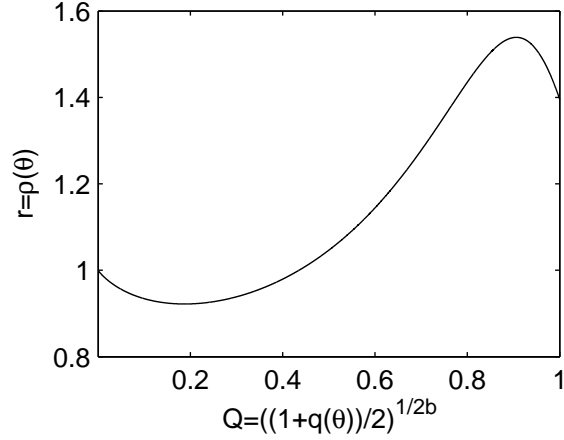,height=6cm} 
\end{center}
\end{figure}

Although \gzit{e.asEOS} is a complete and fully adequate description of 
the thermodynamic state space close to the critical point, we now 
discuss whether we can also find a smooth representation of the 
$D$-explicit form \gzit{e.critSimple}. 
There is a generally accepted belief that it
``{\em is not possible to write the scaled expression for $D$ as 
an explicit function of\, $\Sigma$ and $\Theta$.  
Such attempts always cause singular behavior of the thermodynamic 
potential in the one-phase region either at $\Sigma =0$ or at 
$\Theta=0$.}''
(This quotes a passage after equation (20) of the recent paper by
\sca{Holten} et al. \cite{HolBAS}, adapted to the present notation.
Essentially the same statement can be found in \sca{Behnejad} et al. 
\cite[Section 10.2.2]{BehSA}. Related issues are discussed in
\sca{Gaunt \& Domb} \cite{GauD} and \sca{Vicentini-Missoni} 
\cite[Section IV]{Vic}.)

The results of the previous section allow us to clarify the extent to 
which this belief is justified, exposing hidden assumptions on which 
it is based. To do this, we choose an arbitrary $a>0$ and define 
\[
b:=\frac{e_0}{e_{-1}}=\frac{1}{\beta\delta}\approx 0.63953,
\]
\[
\rho(\theta):=\sqrt{\theta^2+aW^\crit(\theta)^b},
\]
\[
q(\theta):=\frac{\theta}{\rho(\theta)}.
\]
Clearly, $|q(\theta)|\le 1$, so 
\[
Q(\theta):=\Big(\frac{1+q(\theta)}{2}\Big)^{1/2b}\in[0,1].
\]
Figure \ref{f.criticalQ} shows that
\[
\rho(\theta)=r(Q(\theta))
\]
with a function $r(Q)$ that, by construction, is analytic for 
$Q\in{]0,1[}$. At the boundary, $r(Q)$ is continuously differentiable 
but has higher order singularities, with
\[
r(\delta)=1-c\delta+O(\delta^{2b}),~~
r(1-\delta)=\tmax+c'\delta +O(\delta^{1/b})      \for \delta\downto 0.
\]
This follows since with appropriate positive constants $c_i$, we have 
for $\theta=-1+\eps$ and $\eps\downto 0$,
\[
W^\crit(\theta)=c_1\eps^2+O(\eps^2),~~~
\rho=1-\eps+O(\eps^{2b}),~~~
q=-1+O(\eps^{2b}),~~Q=O(\eps),
\]
and for $\theta=\tmax-\eps$ and $\eps\downto 0$,
\[
W^\crit(\theta)=c_2\eps+O(\eps^2),~~~
\rho=\tmax+c_3\eps^b-\eps+O(\eps^{1+b}),~~~
q=1-O(\eps^{b}),~~Q=1-O(\eps^{b}).
\]
Using \gzit{e.cooNew} and \gzit{e.asEOS}, we find
\lbeq{e.R}
R:=\sqrt{\Theta^2+a|\Sigma|^{2b}}=D^{e_0}\rho(\theta),
\eeq
\[
\Theta=R q(\theta),~~~
\]
so that in view of \gzit{e.R},
\lbeq{e.DRQ}
D=S^\crit(\Sigma,\Theta):=\Big(\frac{R}{r(\Theta/R)}\Big)^{1/e_0}.
\eeq
Thus we have represented $D$ explicitly as a function of $\Theta$ and 
$\Sigma$, as in \gzit{e.critSimple}. 

By construction, $S^\crit(\Sigma,\Theta)$ is analytic except possibly
at $\Theta/R\in\{0,1\}$, which corresponds to $\Sigma=0$.
However, a similar argument shows that the formula \gzit{e.DRQ} also 
holds with 
\[
\wt\rho(\theta):=\Big(W^\crit(\theta)+a|\theta|^{c}\Big)^{1/c},~~~
\wt R:=(\Sigma^2+a|\Theta|^{c})^{1/c}
\]
in place of $R$ and $\rho(\theta)$, where 
$c:=2e_{-1}/e_0\approx 3.12728$. The resulting alternative
expression shows that $S^\crit(\Sigma,\Theta)$ (which must obviously be
independent of the way it is represented) is analytic except possibly
at $\Theta$. We conclude that $S^\crit(\Sigma,\Theta)$ is in fact 
analytic throughout the one-phase region.\footnote{
Thus the nonanalyticity in  $R$ is completely cancelled by that in 
$r(Q)$. Such a complete cancellation is not necessarily the case for 
other critical systems. For example, the  simplified but exactly 
solvable 1-dimensional model of a fluid with a critical point discussed 
by \sca{Fisher \& Felderhof} \cite{FisFa,FisFb} -- whose 
$\Sigma$-explicit EOS, given in footnote ${}^{\ref{f.FF}}$, is slightly 
different from that discussed here --, has a continuous phase 
transition for $\Theta=0>\Sigma$, corresponding to nonanalytic behavior 
of $S^\crit(\Sigma,\Theta)$ at this locus.
(They call $\Theta=0$ the limit locus and $\Sigma=0$ the vapor-pressure 
curve.)
} 

However, when $r(Q)$ is approximated by a polynomial or another 
analytic function, it has no longer the correct nonanalytic structure 
at the boundary to cancel the nonanalyticity in $R$, and the resulting 
approximation to $S^\crit(\Sigma,\Theta)$ becomes nonanalytic at 
$\Sigma=0$. 

As \gzit{e.DRQ} shows, it is possible to write $D$ as an analytic and 
explicit function of $\Sigma$ and $\Theta$. In this formulation, the 
lack of analyticity in the one-phase region alluded to in the  
statement quoted above appears only when one approximates the exact 
$r(Q)$ by a polynomial, and can therefore, in principle, be made 
numerically arbitrarily small. But since a highly accurate 
approximation of nonanalytic functions by polynomials needs a very 
high polynomial degree, we do not recommend the $D$-explicit form for 
numerical work.

The $\Sigma$-explicit EOS does not share these approximation 
difficulties, since $C(\theta)$ is analytic in the closed interval 
$[-1,\tmax]$ and easy to approximate by low degree polynomials.

\section{A global $\Sigma$-explicit scaling equation of state}
\label{s.explicit} 

While it is unclear how to extend Schofield's generalized polar 
coordinates approach to the case further away from the critical point, 
when the scaling corrections cannot be neglected, the $\Sigma$-explicit 
form has an immediate generalization.

The new formulation provides us with very useful additional flexibility.
Indeed, if we rearrange the equation \gzit{e.iEOS0} into any 
algebraically equivalent form, the content of the equation remains 
unchanged. This means that many different functions $s$ express the same
constraint on the thermodynamic state space, and we can look for a
normal form that is simpler and restores the uniqueness. 
It turns out that for states close to a critical point belonging to the 
Ising universality class, we can solve \gzit{e.iEOS0} uniquely for 
$\Sigma^2$. This\footnote{\label{f.FF}
The exactly solvable model fluid of \sca{Fisher \& Felderhof} 
\cite{FisFb} possesses close to the critical point an asymptotic 
scaling EOS of the closely related form 
$\Sigma=D^{e_{-1}}X(D^{-e_0}\Theta)$. The variables of their case (Aii) 
are related to the present notation by\vspace{-0.2cm}
\[
\xi=\Sigma,~~\theta=\Theta,~~\omega=D,~~~~T_c=1+e_{-1},~~\sigma=e_0.
\]
} 
allows us to cast the equation 
as a \bfi{scaling EOS in normal form}
\lbeq{e.sig2}
\Sigma^2=D^{2e_{-1}}W(D^{-e_0}\Theta,
                D^{-e_1}I_1,D^{-e_2}I_2,\ldots), ~~~D>0,
\eeq
valid except at critical points, where $D=\Sigma=\Theta=0$.
Note that \gzit{e.sig2} is explicit in $\Sigma^2$ rather than in $D$. 
The squared scaling field reflects the mirror symmetry of the Ising 
model and allows for phase transitions at $\Sigma=0$.  

Generalizing from the asymptotic situation close to a critical point, 
discussed above, the scaling EOS holds for the Ising universality class
in the $\Sigma$-explicit normal form \gzit{e.sig2} with 
\lbeq{e.C}
W(\theta,\iota_1,\iota_2,\ldots)
=\Big(L(\iota_1,\iota_2,\ldots)-\theta\Big)
          C(\theta,\iota_1,\iota_2,\ldots)^2,
\eeq
with universal nonnegative, analytic functions $L$ and $C$ satisfying
\[
L(0,0,\ldots)=\tmax,~~~
C(\theta,0,0,\ldots)=C^\crit(\theta).
\]
As a consequence, the coexistence manifold $\Sigma=0>\Theta$ is now 
characterized by the equation $C(\theta,\iota_1,\iota_2,\ldots)=0$. 
The \bfi{equation for the coexistence manifold} therefore takes the form
\lbeq{e.coex}
C(D^{-e_0}\Theta,D^{-e_1}I_1,D^{-e_2}I_2,\ldots)=0.
\eeq
\at{extrapolation to the metastable case + refs}

The novelty is that, in this form, $W$ is a universal 
{\em analytic} function of its arguments. Since  $D>0$ off the critical 
manifold, the powers of $D$ appearing in \gzit{e.sig2} introduce no 
singularities in the thermodynamic state space. Thus away from the 
critical manifold, $\Sigma^2$ is also an analytic function of $D$, 
$\Theta$, and the $I_k$. The nonanalyticity shows only upon trying to 
solve for $D$ (or variables on which $D$ depends).

The derivation of \gzit{e.sig2} and \gzit{e.C} is valid only close to 
a critical point, hence \gzit{e.sig2} is guaranteed to hold only there. 
However, if $B$ and $C$ are analytic functions without singularities 
(an assumption similar to Griffiths analyticity; cf. 
\sca{Griffiths} \cite{Gri.an}) then analytic continuation implies its 
global validity.

It would be interesting to find explicit numerical expressions for
the universal functions $L$ and $C$. However, only limited information 
is available at present about the numerical details of the corrections 
to scaling; see, e.g.,
\sca{Butera \& Comi} \cite{ButC},
\sca{Campostrini} et al. \cite{CamPRV9}, 
\sca{Hasenbusch} \cite{Has.Fin},
\sca{Zhong \& Barmatz} \cite{ZhoB}.

\section{Phenomenological scaling models for fluid mixtures}
\label{s.phenScal} 

To apply the general results to specific fluid mixtures, one needs
to take the substance-specific information into account. It enters 
through the way the scaling fields $\Theta=\Theta(T,\mu)$, 
$\Sigma=\Sigma(T,\mu)$, $I_k=I_k(T,\mu)$, and $D=D(P,T,\mu)$ are 
expressed in terms of the basic thermodynamic variables. 
Apart from the smoothness requirement, the general theory is silent 
about their form, which varies from system to system.
The literature contains a large variety of phenomenological scaling 
models that provide reasonable formulas for the EOS of fluid mixtures 
that can be related to experimental information.

In the first paper modeling (binary) mixtures with correct scaling 
properties close to the critical point, \sca{Leung \& Griffiths} 
\cite{LeuG} express everything in terms of force field variables.
For industrial applications, multicomponent formulations in a cubic 
EOS, Helmholtz or Gibbs free energy framework are desirable, so that 
the composition can be kept constant.
A number of such formulations were presented in the literature; see, 
e.g., \cite{AniGKS,BelKR,EdiAS,JinTS,KisE,KisF,KisR,RaiF}.
However, formulations at fixed composition cannot match exactly the
singularities; indeed \sca{Wheeler \& Griffiths} \cite{WheG}
prove that when some mole fractions are held constant, curves of plait 
points have bounded heat capacity, while the heat capacity at the 
critical point of a pure substance diverges.
A more detailed analysis leads to additional renormalization phenomena 
(\sca{Fisher} \cite{Fis0}). Ignoring these, as often done in these 
formulations, therefore requires additional approximations which may 
result in artifacts very close to the critical point 
(\sca{Kiselev \& Friend} \cite{KisF}). 
Other papers (e.g., \cite{CaiP,CaiQZH}) implement the renormalization 
group approach more directly, resulting in an iterative definition of an
equation of state that in the limit of infinitely many iterations 
satisfies the correct scaling laws.
A thorough discussion of many practically relevant issues is given in 
the surveys by \sca{Anisimov \& Sengers} \cite{AniS} and \sca{Behnejad}
et al. \cite{BehSA}. 

In most models for critical fluids, $D$ is represented in the special 
form
\lbeq{e.Dpi}
D(P,T,\mu)=P_\regu(T,\mu)-P
\eeq
with a smooth function $P_\regu$ of $T$ and $\mu$, and the remaining 
scaling fields $\Sigma$, $\Theta$ and the $I_k$ are independent of 
$P$. 
One characterizes this situation by saying that the resulting scaling 
laws are based on the assumption of \bfi{revised scaling} 
(\sca{Rehr \& Mermin} \cite{RehM}). 
Under these assumptions, \gzit{e.D} gives an EOS expressing the 
pressure\footnote{
In ``magnetic'' language, the pressure is essentially the free energy. 
} 
explicitly as a function of temperature and chemical potential,
\[
P=P_\regu(T,\mu)-\Delta(T,\mu),
\]
where
\[
\Delta(T,\mu)
:= S(\Sigma(T,\mu),\Theta(T,\mu),I_1(T,\mu),I_2(T,\mu),\dots)
\]
is interpreted as a singular \bfi{crossover term} correcting the 
classical mean field contribution $P_\regu(T,\mu)$ to the pressure in
order to ensure the correct critical behavior.

However, the analysis of recent experiments suggests that revised 
scaling is not sufficient for some fluids; see the discussion in 
\sca{Bertrand} et al. \cite{BerNA}. 
Revised scaling also appears to be not general enough for theoretical 
reasons; e.g., the scaling fields of the model fluid of 
\sca{Fisher \& Felderhof} \cite{FisFb} respect all standard 
requirements of statistical mechanics but do not satisfy the assumption 
of revised scaling.
The more general \bfi{complete scaling} approach developed by 
\sca{Fisher \& Orkoulas} \cite{FisO,KimFO}, although without a strong 
theoretical foundation,\footnote{
It seems that all current theoretical derivations of the scaling laws
\gzit{e.D}--\gzit{e.scaling} from renormalization group arguments 
apply to fluids only in the context of revised scaling. 
A derivation that provides a theoretical justification for complete 
scaling would be welcome. 
} 
covers the above exceptions from revised scaling. 
In this approach, $\Theta$ and $\Sigma$ are allowed to be 
pressure-dependent, so that pressure, temperature and chemical potential
enter symmetrically into the relevant scaling fields.
In particular, \sca{Bertrand} et al. \cite{BerNA} suggest that current 
experimental critical data are consistent with keeping $\Theta$ 
pressure-independent, while $\Sigma$ must slightly depend on $P$ to 
correctly account for the so-called Yang--Yang anomaly \cite{FisO}.

Note that once $\Sigma$ or $\Theta$ is $P$-dependent, there is no 
particular reason why one should abolish the asymmetry in pressure, 
temperature and chemical potential while keeping an asymmetry in the 
scaling fields by restricting $D$ to the special form \gzit{e.Dpi}. 
Thus we propose to allow $D(P,T,\mu)$ to have an arbitrary smooth 
$P$-dependence. This leads to a more complete symmetry between the 
scaling fields and more flexibility in the adaptation to experimental 
data.

\section{A multiphase critical equation of state}\label{s.multiphase}

Most previous critical EOS are limited in several different ways:

$\bullet$
Frequently, the thermal scaling field $\Theta$ is taken to be linear in 
the temperature. However, linearity in $T$ limits the EOS to a narrow 
range of temperatures. Scaling fields with a more favorable temperature 
dependence such as one linear in $T^{-1}$ or $\tanh(T_0/T)$ 
extrapolate much better to the high temperature regime; see, e.g., 
{\sc Lundow \& Campbell} \cite{LunC}.

$\bullet$ 
Almost all studies attempting to go beyond the immediate neighborhood 
of the critical point work in the simplified setting of revised scaling,
which does not account for all observable fluid behavior.
The only previous EOS not restricted to revised scaling is the 
crossover EOS of \sca{Bakhshandeh \& Behnejad} \cite{BakB,BakB2}, which 
employs complete scaling, with scaling fields linear in $P$, $T$, and 
$\mu$.

$\bullet$ 
All noniterative equations of state with correct critical scaling are 
currently based on an implicit representation in terms of a Schofield 
type parameterization.

$\bullet$ 
Most papers discuss the two-phase case only. 
The only exception is \sca{Rainwater} \cite{Rai}, who attempts to cover 
vapor-liquid-liquid equilibrium. 
The main reason for this lack of generality seems to be that (as 
Rainwater's paper shows) a crossover mechanism in terms of a 
Schofield type parameterization is very difficult to extend to the 
multiphase case, since the Schofield parameters have no clear meaning 
far from the critical point and tend to introduce unphysical artifacts
such as two coexisting vapor phases. 

To construct an EOS suitable for the description of multiple phases 
even close to a critical point, we partition the phases into groups $g$ 
completely separated by a phase space boundary, while the phases within 
each group may be connected with each other by paths in phase space not 
crossing the coexistence manifold. 
One of these groups consists of all fluid phases; for solid phases, the 
groups may consist of a single phase only or (as, e.g., for 
$\beta$-brass, cf. \sca{Lamers \& Schweika} \cite{LamS}) may contain 
several phases related by a critical point.
The phases within each group are described by common, substance-specific
scaling fields $\Sigma_g(P,T,\mu)$, $\Theta_g(P,T,\mu)$, and 
$D_g(P,T,\mu)$, one for each phase group $g$. 
According to the results in Section \ref{s.explicit}, the critical 
behavior\footnote{\label{f.multicrit}
with the possible exception of multicritical points; 
cf. footnote ${}^{\ref{f.tricrit}}$.
However, tricritical points have classical critical exponents 
(\sca{Kortman} \cite{Kor}, \sca{Riedel \& Wegner} \cite{RieW}), so they 
can probably be modelled (as for van der Waals fluids) by the analytic 
nonlinearities in the EOS.
Only logarithmic corrections to scaling (\sca{Stephen} et al. 
\cite{SteAS}, \sca{Wegner \& Riedel} \cite{WegR}) are neglected in such 
models, an approximation probably adequate in most applications.
} 
is correctly modelled if the EOS for each phase group takes 
the form
\lbeq{e.critEOS}
s(D_g^{-2e_{-1}}\Sigma_g^2,D_g^{-e_0}\Theta_g,
                    D_g^{-e_1}I_{g1},D_g^{-e_2}I_{g2},\ldots)=1
\eeq
with a universal, substance-independent function $s$.

\gzit{e.critEOS} is a nonlinear equation repating $P$, $T$, and $\mu$. 
Standard thermodynamic stability considerations imply (cf. 
\sca{Neumaier} \cite{Neu.critEOS}) that, for given $T$ and $\mu$, the 
stable phase is determined by finding all solutions $P$ of the equations
\gzit{e.critEOS} for all phase groups $g$, and taking the one with 
largest $P$. In case of ties, several phases from different phase 
groups coexist.
The Gibbs phase rule for the maximal number of coexistent phases is an 
automatic con sequence of this stability rule.

In particular, \gzit{e.critEOS} is the first EOS that correctly models 
both critical behavior and \bfi{vapor-liquid-liquid equilibrium} (VLLE).
The vapor-liquid equilibrium is described by a coexistence 
equation relating $T$ and $\mu$, which we may write conceptually in 
the form $T=T_{VL}(\mu)$. The degree of freedom lost by enforcing
this coexistence relation reappears as an order parameter specifying 
the relative proportion of the vapor and liquid phases. Similarly,
liquid-liquid equilibrium is described by a coexistence equation 
$T=T_{LL}(\mu)$, and an order parameter specifying the relative 
proportion of the two liquid phases. When the two coexistence surfaces 
meet, i.e., if $T=T_{VL}(\mu)=T_{LL}(\mu)$, we have exchanged two 
lost degrees of freedom by two order parameters specifying the relative 
proportion of the vapor and the two liquid phases. 
Now both the VL and the LL branch of the coexistence manifold must 
satisfy the equation $\Sigma(P,T,\mu)=0$. This is possible only if 
$\Sigma$ is nonlinear; cf. \sca{Anisimov} et al. \cite{AniGKS}.
Already allowing $\Sigma(P,T,\mu)$ to be quadratic in $P$ suffices.
However, in revised scaling models, $\Sigma$ is independent of $P$, 
and $P$ is uniquely determined by $T$ and $\mu$.
This implies that in revised scaling only two simultaneous phases 
within the same phase group are possible, which excludes VLLE.

Returning to the general case, the critical exponents and the universal 
functions $s$ in \gzit{e.critEOS} are independent both of the phase 
group and of the particular mixture. 
The freedom remaining, namely the form of the analytic expressions 
for the scaling fields, must be determined from the known the behavior 
at low density (virial equation of state) and any experimental 
information available.
For phenomenological purposes we may choose all scaling fields as low 
degree polynomials or rational functions of $T$, $P$, and $\mu$ (or, 
as argued in \sca{Neumaier} \cite{Neu.critEOS}, of a reduced 
temperature, a reduced pressure, and reduced activities), and fit the 
coefficients to match experimental data. Simple restrictions discussed 
in \cite{Neu.critEOS} guarantee that, at low densities, one can deduce 
a virial equation of state with the correct multi-component structure.

To exploit the available freedom without incurring artifacts due to 
excessive parameter sensitivity, fitting procedures may make use of all 
techniques available for the construction of modern, accurate 
multiparameter equations of state, as reviewed, e.g., in 
\sca{Lemmon \& Span} \cite{LemS}.
It may be expected that the result will be a 
\bfi{global multi-component EOS} that, by their very 
form, automatically has the following properties:\\
(i) Close to critical points, plait points, and consolute points, the 
correct universality and scaling behavior is guaranteed.\\
(ii) At low densities, one can deduce a virial equation of state with 
the correct multi-component structure.\\
(iii) The Gibbs phase rule holds.\\
(iv) The equation of state makes globally sense, for arbitrary 
thermodynamic conditions in which only the phases modelled are present.

For binary mixtures, a global crossover EOS with correct critical 
scaling and a correct low density limit was 
first derived by \sca{Kiselev \& Friend} \cite{KisF}, using the 
Schofield parameterization; see also \sca{Kiselev \& Ely} 
\cite{KisE.cr}. The present approach is more general, needs no 
parameterization, and works for arbitrarily many components and phases.

\section{Conclusion} \label{s.conclusion}

Starting from generalized polar coordinates slightly more general than 
those used in the traditional Schofield-type parameterizations, we
were able to choose the coordinates in such a way that the
equation of state for the Ising universality class could be 
written in a parameter-free form. 
The resulting EOS expresses the square of the strong scaling field 
$\Sigma$ as an explicit function 
$\Sigma^2=D^{2e_{-1}}W(D^{-e_0}\Theta,D^{-e_1}I_1,\ldots)$ of the 
dependent scaling field $D>0$, the thermal scaling field $\Theta$, and
scaling correction fields $I_1$, $I_2$, \dots, with a smooth, 
universal function $W(\theta,\iota_1,\ldots)$ and the 
universal critical exponents $e_{-1}=\delta/(\delta+1)$, 
$e_0=1/(2-\alpha)$, $e_1=-e_0\Delta_1$, \dots .

We deduced from the general form of the new EOS that -- contrary to an 
in the past generally accepted belief -- the dependent scaling field 
can be written as an explicit function of the relevant scaling fields 
without causing strongly singular behavior of the thermodynamic 
potential in the one-phase region. 

With exact models, all parameterizations would lead to the same 
universal functions. However, different choices for the coordinatizing 
functions in the literature produce numerically very different 
parameterizations, and since all available models are approximate only, 
the universal functions computed from different models are slightly 
different. The EOS presented is the first one that only uses universal 
information (which encodes the underlying physics).
A numerical expression for $W$ was derived, valid close to 
critical points. 

We indicated how to use the new EOS to model multiphase fluid 
mixtures, in particular for vapor-liquid-liquid equilibrium (VLLE)
where the traditional revised scaling approach fails.
This is an important first step for the construction of industrial 
strength thermodynamic models for critical fluids and fluid mixtures of 
practical importance that are accurate over the whole experimentally 
accessible thermodynamic phase space.

\appendix

\section{Universality in generalized polar coordinates}
\label{s.uniGen}

\def\xmax{\xi_{\max}}
\def\tmax{\theta_{\max}}
\def\coex{{\fns{coex}}}

Because of the freedom in normalizing the scaling fields, a generalized 
polar coordinate system is universal only up to two positive scaling 
constants. 
Rescaling the basic fields according to the renormalization group 
transformation \gzit{e.scTrans} with arbitrary $\lambda>0$ does not 
change the relations between the fields $D$, $\Sigma$, $\Theta$, and 
$\Omega$. Considerations of scaling may therefore assume without loss 
of generality that one of these fields is unchanged, which we take to 
be $\Theta$. 
Because of the definition of $\Omega$, this leaves two scaling degrees 
of freedom, given by the family of \bfi{scale transformations}
\lbeq{e.scalTrans}
\ol D=mhD,~~~
\ol \Sigma=h\Sigma,~~~
\ol\Theta=\Theta,~~~
\ol \Omega=m\Omega.
\eeq
The corresponding transformation of the parameterizing functions 
$c_D$, $c_\Theta$, $c_\Sigma$, and $c_\Omega$ is
\[
c_{\ol D}=mhc_D,~~~
c_{\ol \Sigma}=hc_\Sigma,~~~
c_{\ol\Theta}=c_\Theta,~~~
c_{\ol \Omega}=mc_\Omega,
\]
Using the traditional critical exponents 
\gzit{e.betaDelta}, we find that the fractions
\lbeq{e.FGK}
F:=|\Omega|^{-\delta}\Sigma,~~~
G:=|\Theta|^{-\beta\delta}\Sigma,~~~
K:=|\Theta|^{-\beta}\Omega,~~~
\eeq
\lbeq{e.XYZ}
X:=|\Omega|^{-1/\beta}\Theta,~~~
Y:=|\Sigma|^{-1/\beta\delta}\Theta,~~~
Z:=|\Sigma|^{-1/\delta}\Omega,~~~
\eeq
do not depend on the radial coordinate $r$, hence they are invariant 
under renormalization transformations. They scale by constant factors 
under scale transformations \gzit{e.scalTrans}.

\begin{figure}[t!]
\caption{The functions $f(\xi)$, $g(\xi)$, $k(\xi)$, $x(\xi)$, $y(\xi)$,
$z(\xi)$ for the Ising universality class in the $\Sigma$-explicit 
representation, where $\xi=\tmax-\theta$. The scaling constants of this
representation are $F_0=1.162123$ and $X_0=0.617917$.}
\label{f.criticalFG}
\begin{center}
    \psfig{figure=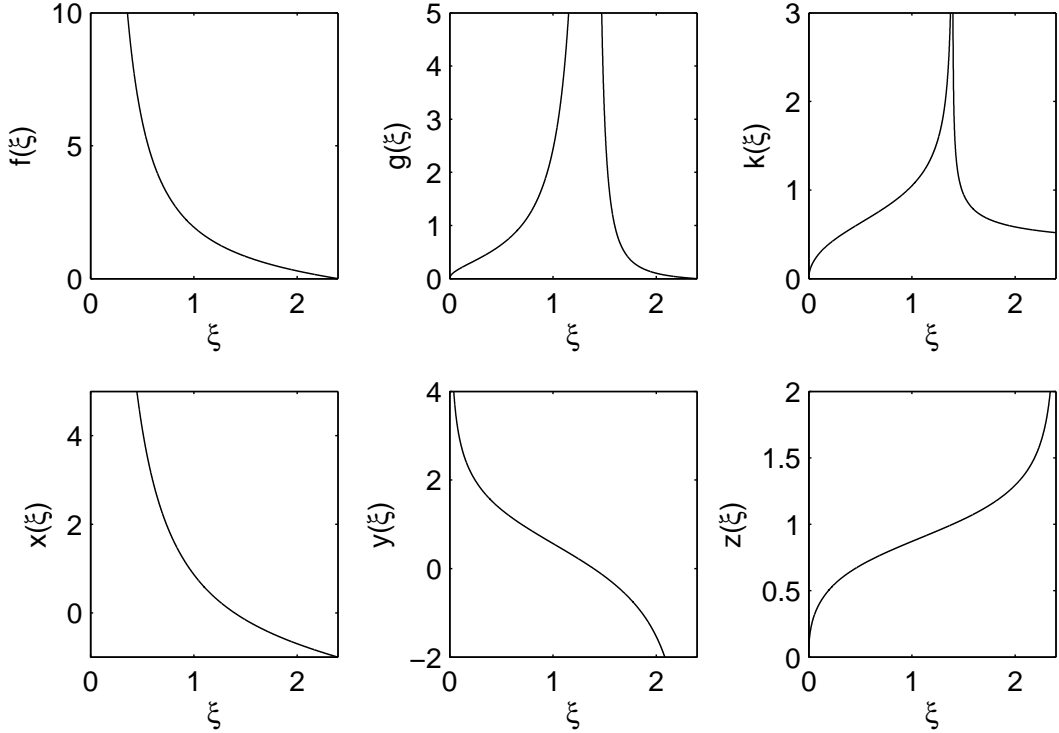,height=10cm} 
\end{center}
\end{figure}

In an arbitrary generalized polar coordinate representation, we 
introduce the functions
\lbeq{e.tso}
\theta(\xi):=c_\Theta(\xi),~~~
\sigma(\xi):=\sqrt{\xi} c_\Sigma(\xi),~~~
\omega(\xi):=\sqrt{\xi} c_\Omega(\xi).
\eeq
For models from the Ising universality class, $\sigma(\xi)$ and 
$\omega(\xi)$ are positive for $0<\xi<\xmax$,  hence
\lbeq{e.sign}
\sign \Omega=\sign \Sigma.
\eeq
This implies that, without loss of generality, we may restrict 
attention to states with $\Sigma\ge 0$. For such states, we find that 
the above fractions are functions
\[
F(\xi):=|\omega(\xi)|^{-\delta}\sigma(\xi)\ge 0,~~~
G(\xi):=|\theta(\xi)|^{-\beta\delta}\sigma(\xi)\ge 0,~~~
K(\xi):=|\theta(\xi)|^{-\beta}\omega(\xi)\ge 0,~~~
\]
\[
X(\xi):=|\omega(\xi)|^{-1/\beta}\theta(\xi),~~~
Y(\xi):=|\sigma(\xi)|^{-1/\beta\delta}\theta(\xi),~~~
Z(\xi):=|\sigma(\xi)|^{-1/\delta}\omega(\xi)\ge 0
\]
of $\xi$ that are analytic for $0<\xi<\xmax$. 
Only two of these functions are independent; expressed through $F$ and 
$X$, the other functions are given by 
\[
G=F|X|^{-\beta\delta},~~~
K=|X|^{-\beta},~~~
Y=XF^{-1/\beta\delta},~~~
Z=F^{-1/\delta}.
\]
The two positive \bfi{scaling constants} 
\lbeq{e.FX0}
F_0:=F(\xi_0),~~~X_0:=-X(\xmax)
\eeq
determine the scale of the generalized polar coordinates. Indeed, it is 
not difficult to see that with the additional scaling constants 
\[
G_0:=F_0X_0^{-\beta\delta},~~~
K_0:=X_0^{-\beta},~~~
Y_0:=X_0F_0^{-1/\beta\delta},~~~
Z_0:=F_0^{-1/\delta},
\]
the quotients
\lbeq{e.fz}
f:= F/F_0,~~~
g:=G/G_0,~~~
k:=K/K_0,~~~
x:=X/X_0,~~~
y:=Y/Y_0,~~~
z:=Z/Z_0
\eeq
satisfy 
\lbeq{e.gfx}
g=fx^{-\beta\delta},~~~k=x^{-\beta},~~~
y=xf^{-1/\beta\delta},~~~z=f^{-1/\delta},
\eeq
and are \bfi{universal}, since they are invariant under both the 
renormalization transformations and the scale transformations 
\gzit{e.scalTrans}. For the $\Sigma$-explicit representation, they are 
drawn in  Figure \ref{f.criticalFG} in dependence on $\xi$.
%
\begin{figure}[t!]
\caption{Universal relations between the asymptotic reduced fractions.
Note that the graphs of $g_+(k)$ and $g_-(k)$ cross each other.}
\label{f.criticalXY}
\begin{center}
    \psfig{figure=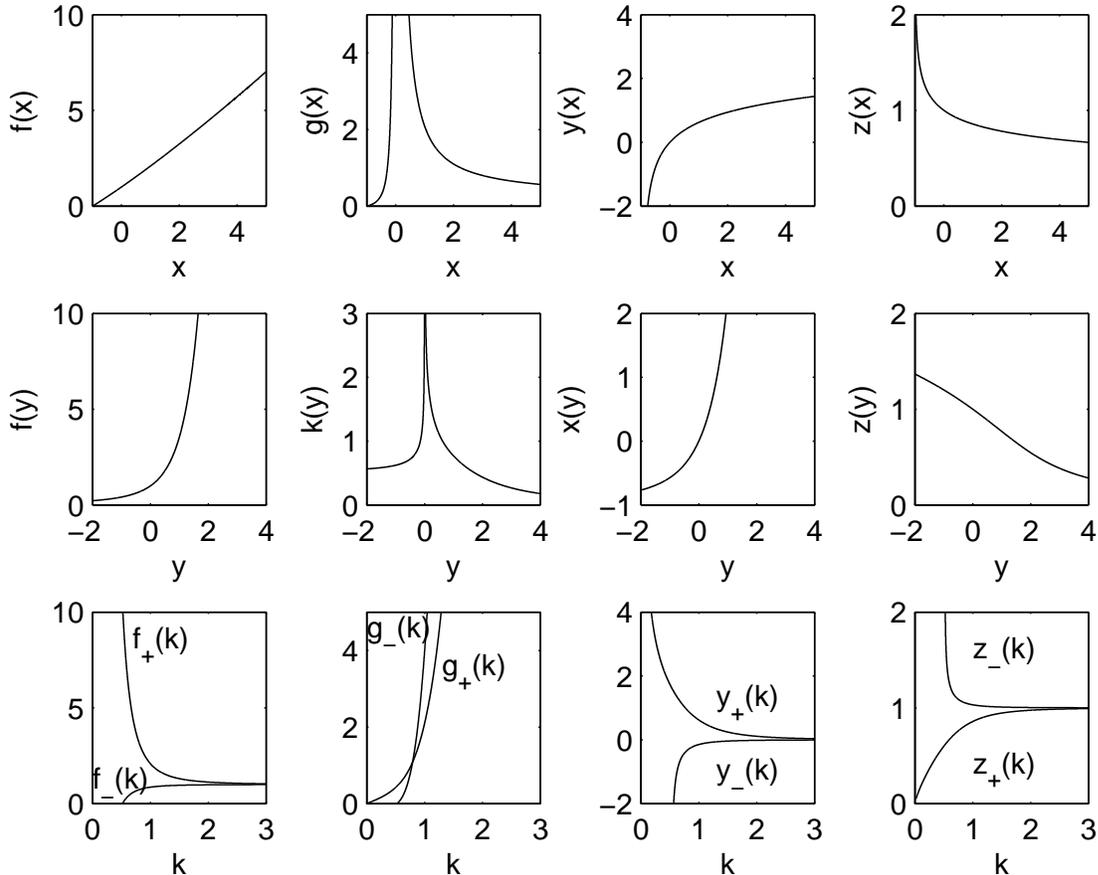,height=12cm} 
\end{center}
\vspace*{-1cm}
\end{figure}
%
Since $x$ and $y$ are strictly monotone functions of $\xi$, we may 
express all quotients as functions of $x \in [-1,\infty]$ or 
$y\in[\infty,\infty]$ rather than $\xi\in [0,\xmax]$. Apart from 
\gzit{e.gfx}, this gives the nontrivial relations
\[
F=F_0f(x),~~~G=G_0g(x),~~~Y=Y_0y(x),~~~Z=Z_0z(x),
\]
\[
F=F_0f(y),~~~K=K_0g(y),~~~X=X_0x(y),~~~Z=Z_0z(y)
\]
(cf. Figure \ref{f.criticalXY}), 
involving universal analytic functions (denoted with a slight abuse of 
notation by) $f(x)$, $g(x)$, $y(x)$, $z(x)$ and $f(y)$, $g(y)$, $x(y)$, 
$z(y)$, respectively, expressing in different ways the same EOS relating
$\Omega$, $\Sigma$ and $\Theta$.
For example, $F=F_0f(x)$ becomes the EOS\footnote{
Since $f(x)$ satisfies the normalization conditions $f(-1)=0$ and 
$f(0)=1$, this shows that $f(x)$ is identical with the universal 
function defined in \sca{Pelissetto \& Vicari} \cite[(1.71)]{PelV}.
} 
\[
\Sigma = F_0 |\Omega|^\delta f(X_0^{-1}|\Omega|^{-1/\beta}\Theta).
\]
To compare more directly with microscopic data, we use the fact that 
$k(\xi)$, which has a singularity at $\xi=\xi_0$, is strictly monotone  
for $\xi<\xi_0$ (the \bfi{high temperature regime} $\Theta>0$) and for 
$\xi>\xi_0$ (the \bfi{low temperature regime} $\Theta<0$). 
Thus, for $\pm\Theta>0$, we may solve $k=k(\xi)$ for $\xi=\xi_\pm(k)$ 
and express all quotients as functions of $k$ rather than $\xi$. 
This gives the trivial relation $x=k^{-1/\beta}$ and the nontrivial 
relations
\[
F=F_0f_\pm(k),~~~G=G_0g_\pm(k),~~~Y=Y_0y_\pm(k),~~~Z=Z_0z_\pm(k)
\]
(cf. Figure \ref{f.criticalXY}), 
involving universal analytic functions $f_\pm(k)$, $g_\pm(k)$, 
$y_\pm(k)$, $z_\pm(k)$. When $\xi$ is small, 
$k(\xi)=\xi^{1/2}(c+O(\xi))$ with a constant $c\ne0$. 
This implies that $\xi_+(k)$ is an even analytic function of $k$ 
vanishing at $k=0$. This implies that we may write
\lbeq{e.gHT}
g_+(k)=g_0F_{HT}(k/k_0) \for \Theta>0,
\eeq
where
\lbeq{e.HT}
F_{HT}(s)=\sum_{n=1}^\infty \frac{r_{2n}}{(2n-1)!} s^{2n-1}
\eeq
is an odd universal function,\footnote{
With $z_0:=k_0^{-1}$ and $F_0^\infty:=g_0^{-1}k_0^\delta$, \gzit{e.gfx} 
implies that $s=k/k_0=z_0x^{-\beta}$ and 
$s^{-\delta}F_{HT}(s)=F_0^\infty f(x)$. Hence $F_{HT}(s)$ is identical 
with the function called $F(z)$ in \sca{Pelissetto \& Vicari} 
\cite[(1.84)]{PelV}. 
} 
which expresses the relation between $\Omega$, $\Sigma$ and $\Theta$ 
in the form of the \bfi{high temperature EOS}
\lbeq{e.MEOS}
\Sigma=a|\Theta|^{\beta\delta} F_{HT}(b|\Theta|^{-\beta}\Omega)
\for \Omega>0
\eeq
with $a=g_0G_0$ and $b=(k_0K_0)^{-1}$.
In \gzit{e.gHT}, $g_0$ and $k_0$ are universal scaling constants, 
chosen such that the coefficients of $F_{HT}(z)$ satisfy the 
normalization conditions
\[
r_2=r_4=1.
\]
Values for the other coefficients $r_{2k}$ of the \bfi{high temperature 
expansion} \gzit{e.HT} have been computed from microscopic models by 
a number of different methods. 
\sca{Butera \& Pernici} \cite[Table X]{ButP} survey these methods and 
gives as presently most accurate values
\lbeq{e.rNum}
r_6=2.061(2),~~~r_8=2.54(4),~~~r_{10}=-15.2(4),~~~r_{12}=45(5),~~~
r_{14}=1400(200).
\eeq
(On the other hand, the universal analytic function $g_-(k)$ is not odd 
since $\xi_-(k)$ lacks the reflection symmetry of $\xi_+(k)$.)

\at{To determine reliable coefficients from 
this underdetermined system, we may invoke the principle of minimal 
sensitivity (\sca{Stevenson} \cite{Ste}), e.g., by making $\tmax$ 
stationary or by enforcing the smallest possible dependence on $r_{14}$.
cf. P/V p. 610f}

\section{Comparison with data from the literature}
\label{s.num}

Different generalized polar coordinate systems simply express these  
universal functions in different parametric forms. With exact models,
all generalized polar coordinate systems would lead to the same 
universal functions. However, since all available models are 
approximate only, the universal functions computed from different models
are slightly different. 
The traditional choice is based on arguments of simplicity, and uses
\lbeq{e.fTrad}
c_\Theta(\xi):=1-\xi,~~~c_\Omega(\xi):=m_0
\eeq 
with a constant $m_0$ determined implicitly by the normalization 
$c_\Sigma(0)=1$. The resulting generalized polar coordinates
have a qualitative geometrical meaning but are devoid of a clear 
physical meaning. (The same holds for the nontraditional choices 
discussed, e.g., in \sca{ Fisher \& Zinn} \cite{FisZ} and 
\sca{Pelissetto \& Vicari} \cite[Section 3.4]{PelV}.)

Indeed, different choices for the two coordinatizing functions may
produce numerically very different parameterizations. Only the implicit 
functions defined by the parameterization \gzit{e.genPolar}
are (apart from scaling factors) universal as only they correspond to
the underlying physics.
For example, the numerical details of the forms reported in 
\sca{Butera \& Pernici} \cite{ButP} 
and \sca{Campostrini} et al. \cite{CamPRV} 
are completely different, although they make very similar physical 
predictions and both start with the same coordinatizing 
functions\footnote{
\sca{Butera \& Pernici} \cite{ButP} use $M=m_0R^\beta\theta$, 
$\tau=R(1-\theta^2)$, and $h=h_0 R^{\beta\delta}\ell(\theta)$. 
This corresponds in the present notation to
\[
\Omega=m_0 r^\beta \phi,~~~\Theta=r(1-\phi^2),~~~
\Sigma=h_0 r^{\beta\delta}\ell(\phi),
\]
consistent with \gzit{e.eTrad}, though their constants $m_0$ and $h_0$ 
amount to a different scaling than that defining our $\Sigma$-explicit
EOS.
\sca{Campostrini} et al. \cite{CamPRV} use essentially the same model
but they write $h(\theta)$ in place of $\ell(\theta)$ and use a 
different value of $m_0$.
} 
\gzit{e.fTrad}. Only the nonuniversal constant $m_0$ in \gzit{e.fTrad} 
is different, \at{find the conversion factor!}
but this is enough to completely change the numerical coefficients:
\sca{Butera \& Pernici} \cite[eq. (24)]{ButP} report the numerical 
approximations $\phi_{\max}=1.1273$ and 
{\small
\lbeq{e.BP}
c_\Sigma(\xi)=1-0.8014(50)\xi+0.00946(30)\xi^2+0.00141(40)\xi^3
+0.00029(10)\xi^4-0.00011(5)\xi^6;
\eeq
}
\sca{Campostrini} et al. \cite[Table V]{CamPRV} report the numerical 
approximations $\xmax=1.37861$ (corresponding to  $\phi_{\max}=1.17414$)
and 
\lbeq{e.CamPRV}
c_\Sigma(\xi)=1-0.736743\xi+0.008904\xi^2-0.000472\xi^3.
\eeq
(For other sources of numerical data, see
\sca{Agayan} et al. \cite{AgaAS},
\sca{Campostrini} et al. \cite{CamPRV}, 
\sca{Guida \& Zinn-Justin} \cite{GuiZ}, 
\sca{Hasenbusch} \cite{Has}.)

We checked that in the range plotted, the critical functions computed 
from the $\Sigma$-explicit EOS with $C(\theta)$ given by 
\gzit{e.Gfit}--\gzit{e.Gfit1}, differ from those computed from the 
(more recent and more accurate) parameterization by 
\sca{Butera \& Pernici} \cite[eq. (24)]{ButP} by a root mean squared 
error of less than $2.5\cdot 10^{-5}$. Indeed, the parameters were
derived by numerical optimization of a measure of discrepancy between
these functions.

\bigskip

\end{document}